\documentclass[fleqn,10pt]{wlscirep}
\usepackage[utf8]{inputenc}
\usepackage[T1]{fontenc}
\usepackage{textcomp}
\usepackage{ulem}
\usepackage[separate-uncertainty = true, multi-part-units=single, range-phrase=--, range-units = single, detect-all]{siunitx}
\DeclareSIUnit\atomic{at.}

\title{Ultra-low threshold cw and pulsed lasing in tensile strained GeSn alloys}

\author[1,2]{Anas Elbaz}
\author[3,*]{Dan Buca}
\author[3,4]{Nils von den Driesch}
\author[1]{Konstantinos Pantzas}
\author[1]{Gilles Patriarche}
\author[1]{Nicolas Zerounian}
\author[1]{Etienne Herth}
\author[1]{Xavier Checoury}
\author[1]{Sébastien Sauvage}
\author[1]{Isabelle Sagnes}
\author[5]{Antonino Foti}
\author[5]{Razvigor Ossikovski}
\author[6]{Jean-Michel Hartmann}
\author[2]{Frédéric Boeuf}
\author[7]{Zoran Ikonic}
\author[8]{Philippe Boucaud}
\author[3,4]{Detlev Grützmacher}
\author[1,*]{Moustafa El Kurdi}
\affil[1]{Center for Nanoscience and Nanotechnology, C2N UMR 9001, CNRS, Université Paris Sud, Université Paris Saclay, 91120 Palaiseau, France}
\affil[2]{STMicroelectronics, Rue Jean Monnet 38054 Crolles, France}
\affil[3]{Peter Grünberg Institute (PGI 9) and JARA-Fundamentals of Future Information Technologies, Forschungszentrum Juelich, 52428 Juelich, Germany}
\affil[4]{JARA-Institute Green IT, RWTH Aachen, 52062 Aachen, Germany}
\affil[5]{LPICM, CNRS, Ecole Polytechnique, Université Paris-Saclay, 91128 Palaiseau, France}
\affil[6]{CEA, LETI and Univ. Grenoble Alpes, 38054 Grenoble, France}
\affil[7]{Pollard Institute, School of Electronic and Electrical Engineering, University of Leeds, Leeds LS2 9JT, UK}
\affil[8]{Université Côte d'Azur, CNRS, CRHEA, 06560 Valbonne, France}
\affil[*]{corresponding authors: moustafa.el-kurdi@u-psud.fr, d.m.buca@fz-juelich.de}

\begin{abstract}

GeSn alloys are the most promising semiconductors for light emitters entirely based on group IV elements. Alloys containing more than \SI{8}{\atomic\percent} Sn have fundamental direct band-gaps, similar to conventional III-V semiconductors and thus can be employed for light emitting devices. Here, we report on GeSn microdisk lasers encapsulated with a SiN$_x$ stressor layer to produce tensile strain. A \SI{300}{\nano\meter} GeSn layer with \SI{5.4}{\atomic\percent} Sn, which is an indirect band-gap semiconductor as-grown with a compressive strain of \SI{-0.32}{\percent}, is transformed via tensile strain engineering into a truly direct band-gap semiconductor. In this approach the low Sn concentration enables improved defect engineering and the tensile strain delivers a low density of states at the valence band edge, which is the light hole band. Continuous wave (cw) as well as pulsed lasing are observed at very low optical pump powers. Lasers with emission wavelength of \SI{2.5}{\micro\meter} have thresholds as low as \SI{0.8}{\kilo\watt\per\square\centi\meter} for \si{\nano\second}-pulsed excitation, and \SI{1.1}{\kilo\watt\per\square\centi\meter}  under cw excitation. These thresholds are more than two orders of magnitude lower than those  previously reported for bulk GeSn lasers, approaching these values obtained for III-V lasers on Si. The present results demonstrate the feasabiliy and are the guideline for monolithically integrated Si-based laser sources on Si photonics platform.

\end{abstract}
\begin{document}

\flushbottom
\maketitle
\thispagestyle{empty}

\section*{Introduction}
Si-Ge-Sn alloys are a promising, enabling material system for the monolithic integration of both passive and active optoelectronic devices and circuits  on Silicon \cite{Soref2016}. Silicon (Si) photonics presently relies on integration of III-V materials for emitters \cite{roadmap_2016}. Although such an approach has recently demonstrated some impressive progress, it still faces challenges like the wafer throughput, scalability and compatibility with the current Si complementary metal oxide semiconductor (CMOS) technology. The most successful route for laser action within group IV materials nowadays is based on germanium-tin (GeSn) semiconductors. The first demonstration of an optically pumped laser\cite{wirths_lasing_2015} and subsequent developments to improve the performance in respect to threshold and operation temperature  \cite{stange_optically_2016, reboud_optically_2017, Laser_GeSn_Arkansas, Thai:18}, have shown the potential of these group IV materials for achieving Si-based light sources, the final ingredient for completing an all-inclusive nano-photonic CMOS platform. Furthermore, (Si)GeSn materials can help to extend the present Si photonics platform with a much broader application area than only near infrared data communication. In the short-wave to mid infrared region of \SIrange{2}{4}{\micro\meter}, in which GeSn laser emission has been obtained, potential applications including gas sensing for environmental monitoring and industrial process control\cite{Hodgkinson2013}, lab-on-a-chip applications\cite{Sieger2016, Singh2014} or infrared imaging for night vision and hyperspectral imaging\cite{Razeghi_2014} can be envisaged.\\
An increase in Sn content in GeSn alloys modifies the energy of the electronic bands. The band-gap at {$\Gamma$-point, ($E_\Gamma$) reduces faster than the band-gap towards the $L$-valley ($E_L$), leading to a crossover from an indirect to a direct band-gap semiconductor at an Sn concentration of \SI{8}{\atomic\percent}  \cite{wirths_lasing_2015}. The lattice mismatch between Sn-containing alloys and the Ge buffer layer, the typical virtual substrate for their epitaxial growth, generates compressive strain in the grown layer, which counteracts the effect of Sn incorporation, decreasing the directness $\Delta E_{L-\Gamma} = E_L - E_\Gamma$. On the contrary, applying tensile strain will increase the directness. Finding a proper balance between a moderate Sn content to minimize crystal defects and to maintain thermal stability of the GeSn alloy on one hand and making use of tensile strain on the other hand are the keys to bring lasing threshold and operation temperature close to application's requirements.
The mainstream research to increase $\Delta E_{L-\Gamma}$ focuses on high Sn content alloys \cite{Dou2018a,reboud_optically_2017}, obtained by epitaxy of thick strain-relaxed GeSn layers. A large directnesss is obtained, leading to higher temperature operation, although at the expense of steadily increasing laser threshold \cite{fisher_270K}. We have recently theoretically proposed an alternative approach, which is based on  two key ingredients: employing moderate Sn content GeSn alloys, and inducing tensile strain in them \cite{Rainko2018a}. This study indicated that, if a given directness is reached via tensile strain rather than by increasing Sn content, the material can provide a higher net gain. The underlying physics originates in the valence band splitting and lifting up of the light hole, $LH$, band above the heavy hole, $HH$, band. Its lower density of states (DOS) reduces the carrier density required for transparency, hence reduces the lasing threshold, as will be shown below. \\
GeSn alloys with a moderate Sn content offer a couple of advantages from the materials perspective. The epitaxial growth temperature, \SI{375}{\celsius}  compared to below \SI{300}{\celsius} for high Sn content alloys, yields a better crystalline quality and lower defect density. Lattice mismatch, and therefore, the density of misfit dislocations at the GeSn/Ge interface scales with the Sn content\cite{Gencarelli15032013}. Both types of defects strongly influence the carrier recombination dynamics \cite{Dou2018, StangeMQWlaser} and contribute to the high pumping levels necessary to reach lasing. Accordingly, laser thresholds of \SIrange{100}{300}{\kilo\watt\per\square\centi\meter} were reported at \SI{20}{\kelvin} for GeSn lasers with \SIrange{12}{14}{\atomic\percent} Sn\cite{stange_optically_2016}, while $\sim$\si{\mega\watt\per\square\centi\meter} values are required for very high Sn content alloys (\SI{>20}{\atomic\percent}) above \SI{230}{\kelvin}\cite{Thai:18, Dou2018a}. \\
The described material advantages and the underlying physics should then be combined with the technology able to induce tensile strain in GeSn alloys.  Strain engineering is a mature Si technology employed to modify the electronic band structure of semiconductors\cite{Minamisawa2012}. In pure Ge, \SI{1.7}{\percent} biaxial\cite{el_kurdi_direct_2016, virgilio-rad-2013} or \SI{4.5}{\percent} uniaxial \cite{Bao2017, Suess2013, ArmandPilon_2019} strain is required to energetically align the $L$- and $\Gamma$-valleys ($\Delta E_{L-\Gamma} = 0$), i.e. to reach crossover from an indirect to a direct semiconductor. Even higher strain is required to achieve the mandatory directness, $\Delta E_{L-\Gamma}$ \SI{>150}{\milli\electronvolt}, for room temperature operation of a strained Ge laser. Although such high levels of tensile strain are technologically possible \cite{ghrib_control_2012, capellini_strain_2013}, they are challenging in a laser device geometry. Depending on Sn content, which can be chosen in the range of \SIrange{5}{8}{\atomic\percent}, significantly lower values of tensile strain are needed in GeSn alloys to achieve a sufficient directness \cite{Rainko2018a}. However, in order to obtain laser with low pumping threshold, the impact of the defective GeSn/Ge interface region has to be removed. It is responsible for a considerable part of non-radiative recombination. The growth of hetero- and quantum well structures appears to be a suitable technology to separate the gain material from the defective interface \cite{StangeMQWlaser}. Here, we focus on the impact of tensile strain and, for simplicity, we are using bulk GeSn layers. To remove the defective interface, the layer transfer method is applied, as used to realize GeSn on insulator (GeSnOI) structures. Due to the transfer, the defective part is at the top surface of the GeSn layer and can be easily removed by etching.\\
All GeSn lasers reported in the literature so far operate only under pulsed excitation, although continuous wave (cw) lasing is the key milestone required for accessing the full potential of GeSn for technologically useful optical devices.\\
In this work, we demonstrate both cw and pulsed lasing using microdisk cavities fabricated from initially indirect band-gap Ge$_{0.946}$Sn$_{0.054}$ alloys. Tensile strain of \SI{1.4}{\percent} is applied \textit{ex-situ} to the GeSn layer, using all-around SiN$_x$ stressors. The method relies on formation of GeSnOI by wafer bonding and layer transfer, followed by under-etching so that the final cavity is supported by a metallic post, here Al, that acts as a heat sink. The combination of these factors, i.e. strain engineering, bulk defect density reduction, and improved heat removal, allows the demonstration of lasing in tensile strained GeSn with record low thresholds of \SI{0.8}{\kilo\watt\per\square\centi\meter} and \SI{1.1}{\kilo\watt\per\square\centi\meter} at \SI{25}{\kelvin} in pulsed and cw operation regime, respectively. The threshold is two orders of magnitude lower than previously reported for GeSn pulsed lasers, while no reports of cw lasing in GeSn are available.

\section*{Results}
Material growth, characterization and cavity patterning. The GeSn layers were grown on Ge virtual substrates (Ge-VS) on \SI{200}{\milli\meter} Si(100) wafers\cite{Hartmann2009}, via reactive gas source epitaxy using an AIXTRON TRICENT \textsuperscript{\textregistered} reactor\cite{vondenDriesch2015}. Digermane (Ge$_2$H$_6$) and tin tetrachloride (SnCl$_4$) were used as precursors for elementary Ge and Sn, respectively. GeSn layers with thicknesses of \SI{300}{\nano\meter} and Sn content of \SI{5.4}{\atomic\percent} were grown at \SI{375}{\celsius}. The layers are partially strain relaxed, with a residual compressive strain of \SI{-0.32}{\percent} as measured by X-Ray diffraction (Figure~\ref{fig:PL}a). Details on material characterization can be found in the supplementary information (SI).\\
The GeSn layers were processed into microdisk cavities with all-around SiN$_{x}$ stressor. The technology flow, developed for pure Ge\cite{ghrib_all-around_2015} and adapted for GeSn thermal budget limitations\cite{zaumseil_2018}, is presented in the SI. Yet, some key aspects are shown in Figure~\ref{fig:PL}b. The stressor layer is a \SI{350}{\nano\meter} thick SiN$_{x}$ layer with an intrinsic stress of \SI{-1.9}{GPa}. An additional Al metal layer was added to the layer stack, in order to reduce the heating during optical pumping \cite{Elbaz:18, elbaz_germanium_2018}. This is particularly important, since alloying of Ge with Sn strongly decreases the thermal conductivity of the alloy \cite{uchida_2015}. After GeSn bonding and removal of the donor wafer, the top \SI{40}{\nano\meter} of the GeSn layer, containing the defective GeSn/Ge interface, is also removed. It was shown that this dense misfit network strongly reduces the photoluminescence of the layers at the onset of strain relaxation \cite{pezzoliACSphotonics}. Up to this step the compressive strain in the GeSn layer is preserved (Figure~\ref{fig:PL}c,d). Tensile strain, coming from the SiN$_{x}$ stressor underneath, is induced only by structuring the GeSn/ SiN$_{x}$ layer stack (Figure~\ref{fig:PL}b). The under-etching process, i.e. selective and local removal of Al, was optimized to maximize the tensile strain in GeSn layer and to allow a wide Al post for heat sinking. Subsequently, the suspended disks are conformally covered by a second \SI{400}{\nano\meter} thick SiN$_{x}$ stressor layer, leading to fully encapsulated GeSn disks, standing on an Al post (Figure\ref{fig:PL}b).This layer transfer technology and processing transforms the initial Ge$_{0.946}$Sn$_{0.054}$ layer with residual compressive strain having an indirect band-gap into a microdisk exhibiting pronounced biaxial tensile strain and consequently a direct band-gap.

\begin{figure}[ht]
\centering
\includegraphics[width=\linewidth]{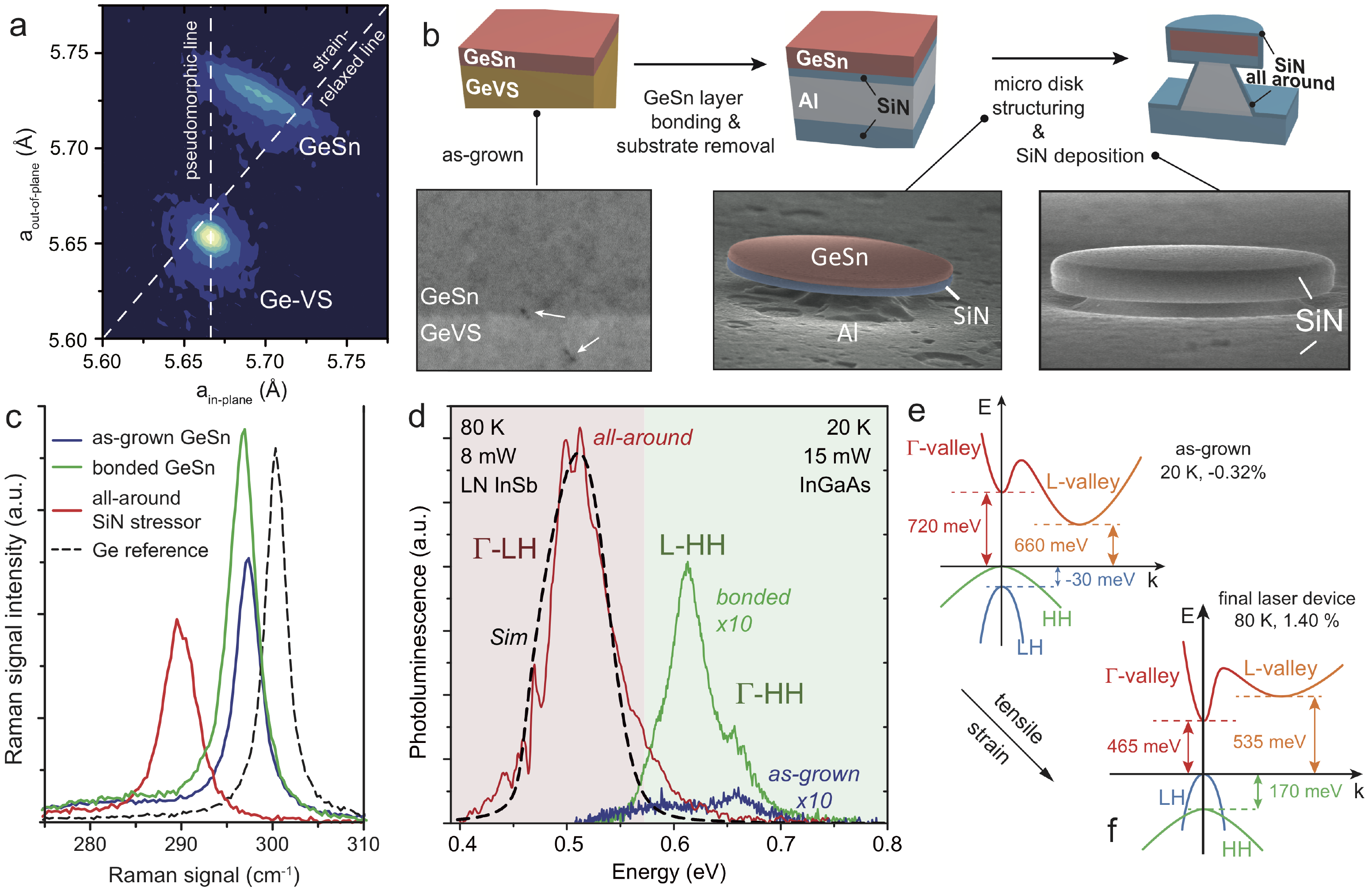}
\caption{\textbf{Structural and optical characterization.}\textbf{(a)} X-Ray diffraction reciprocal space map for quantitative assessment of the layer strain (\SI{-0.32}{\percent}). \textbf{(b)} Sketch of the processing steps. TEM image of the as-grown layer with strain releasing dislocations at the GeSn/Ge-VS interface. SEM images of the under-etched GeSn/SiN$_{x}$ stack and the final laser structure with SiN$_{x}$ all-around. \textbf{(c)} Raman spectra for three cases: as-grown layer, bonded (i.e. transferred) layer and all-around disk of \SI{9}{\micro\meter} diameter. Bulk Ge is used as reference. \textbf{(d)} PL spectra taken at \SI{20}{\kelvin} for the as-grown and bonded GeSn layers (bonded layer has the misfit dislocation network removed), and PL of patterned disk with all-around stressor at \SI{80}{\kelvin}. The dotted line shows the simulated PL spectra of the final structure. The calculated band structure for the \textbf{(e)} indirect band-gap as-grown layer and \textbf{(f)} direct band-gap tensile strained GeSn layer.}
\label{fig:PL}
\end{figure}

Raman spectroscopy was performed to follow the strain evolution during fabrication of GeSn microdisks. The positions of the Ge-Ge vibration modes in GeSn alloys in the as-grown sample (blue line), after bonding onto the host Si substrate (green line), and the final processed microdisk structure are shown in Figure~\ref{fig:PL}c. The Raman modes of the unpatterned transferred layer and the as-grown layer, at \SI{297.3}{\per\centi\meter} and \SI{296.8}{\per\centi\meter}, respectively, are very similar, within the experimental resolution of \SI{0.5}{\per\centi\meter}. Therefore, it can be reasonably assumed that the layer transfer process itself does not change the strain in the GeSn layer. However, after the processing of the final microdisk, a Raman shift of \SI{-10.5}{\per\centi\meter} is detected. Using the equations from Ref.\cite{ChengRamanGeSn2013}, including alloy disorder and strain effects, this Raman shift  corresponds to a built-in biaxial tensile strain of \SI{1.5}{\percent} for the all-around embedded GeSn disk. Note that Raman spectroscopy probes only  the in-plane strain within a small depth below the disk surface, while photoluminescence probes the whole disk volume, giving an average value of the strain, and is directly related to the band structure. \cite{ghrib_all-around_2015}.

Photoluminescence (PL) experiments were conducted to assess the strain-induced band structure changes, as well as the quality improvement of the transferred layer. The PL signal from the as-grown GeSn layer is very weak, due to {\em (i)} the indirect band-gap, with the conduction band energy splitting $\Delta E_{L-\Gamma}=$\SI{-60}{\milli\electronvolt}, and {\em (ii)} the presence of defects at the GeSn/Ge interface. After the layer is transferred, these defects are removed and an increase in PL intensity is observed (Figure~\ref{fig:PL}d). The PL signal is found in the same energy range, thus indicating again that the transferred layer maintains its compressive strain and, therefore, its indirect band-gap character with $L$ and $HH$ as conduction and valence band extrema, respectively. The optical transition at \SI{0.61}{\electronvolt} is attributed to the recombination of electrons in the $L$-valley of the conduction band and holes near the $\Gamma$ point of the $HH$ valence band, thus across the fundamental indirect band-gap. Indirect carrier recombination dominates over the direct transitions, because almost \SI{100}{\percent} of electrons at \SI{20}{\kelvin} are in the $L$-valley. The shoulder of the PL signal around \SI{0.66}{\electronvolt} is assigned to the direct transition, i.e. electrons and heavy holes around the $\Gamma$-point in k-space. Details on optical transition identification can be found in SI. PL spectra in Figure~\ref{fig:PL}d for the as-grown and the transferred GeSn layers are taken under identical conditions with a sensitive InGaAs detector. This is emphasized because this, initially weakly-emitting, layer will become the active laser medium after inducing the tensile strain. \\
The major limitation for the GeSn emission efficiency, the indirect band-gap, is overcome by tensile strain turning it into a direct-gap semiconductor. Since the extended InGaAs detector has a cut-off wavelength of \SI{2.4}{\micro\meter}, another set-up with a nitrogen cooled InSb detector, with a cut-off of \SI{4.8}{\micro\meter}, is used for the fully processed microdisk device. Even though the InSb detector has a lower sensitivity, a strong increase of the integrated PL emission by 2 orders of magnitude compared to  the as-grown layer is measured. The PL signal is strongly red-shifted, showing the peak emission around \SI{0.50}{\electronvolt} which is attributed to the tensile strain of the final structure with the all-around SiN$_{x}$ stressor. The band structure was modeled using \textbf{k$\cdot$p} method, taking \SI{5.4}{\atomic\percent} Sn and a band-gap of \SI{465}{\milli\electronvolt} of the strained film. The best fit to the experimental data in Fig. \ref{fig:PL}d is obtained for biaxial tensile strain of \SI{1.4}{\percent}. This value is slightly smaller than the \SI{1.5}{\percent} obtained from Raman spectroscopy, the discrepancy is attributed to the uncertainty of the parameters used in the two methods. The band structures of the as-grown and tensile strained GeSn are shown in Figure\ref{fig:PL}e and -f, respectively. In the final device, the tensile strain lifts the degeneracy of the $LH$ and $HH$ band, with $LH$ becoming the fundamental valence band. The valence band splitting is $E_{LH}-E_{HH}$=\SI{170}{\milli\electronvolt}. More importantly, in the conduction band the tensile strain shifts the $\Gamma$-valley below the $L$-valley, thus the tensile strained GeSn becomes direct band-gap material with a directness of $\Delta E_{L-\Gamma} = \SI{70}{\milli\electronvolt}$. Consequently, the pronounced enhancement of the PL intensity emission is due to the fundamental direct optical transition. The electrons recombine with $LH$ at the $\Gamma$ point in the center of the Brillouin zone. This transition is labelled as $\Gamma$-$LH$  optical transition.

\begin{figure}[ht]
\centering
\includegraphics[width=\linewidth]{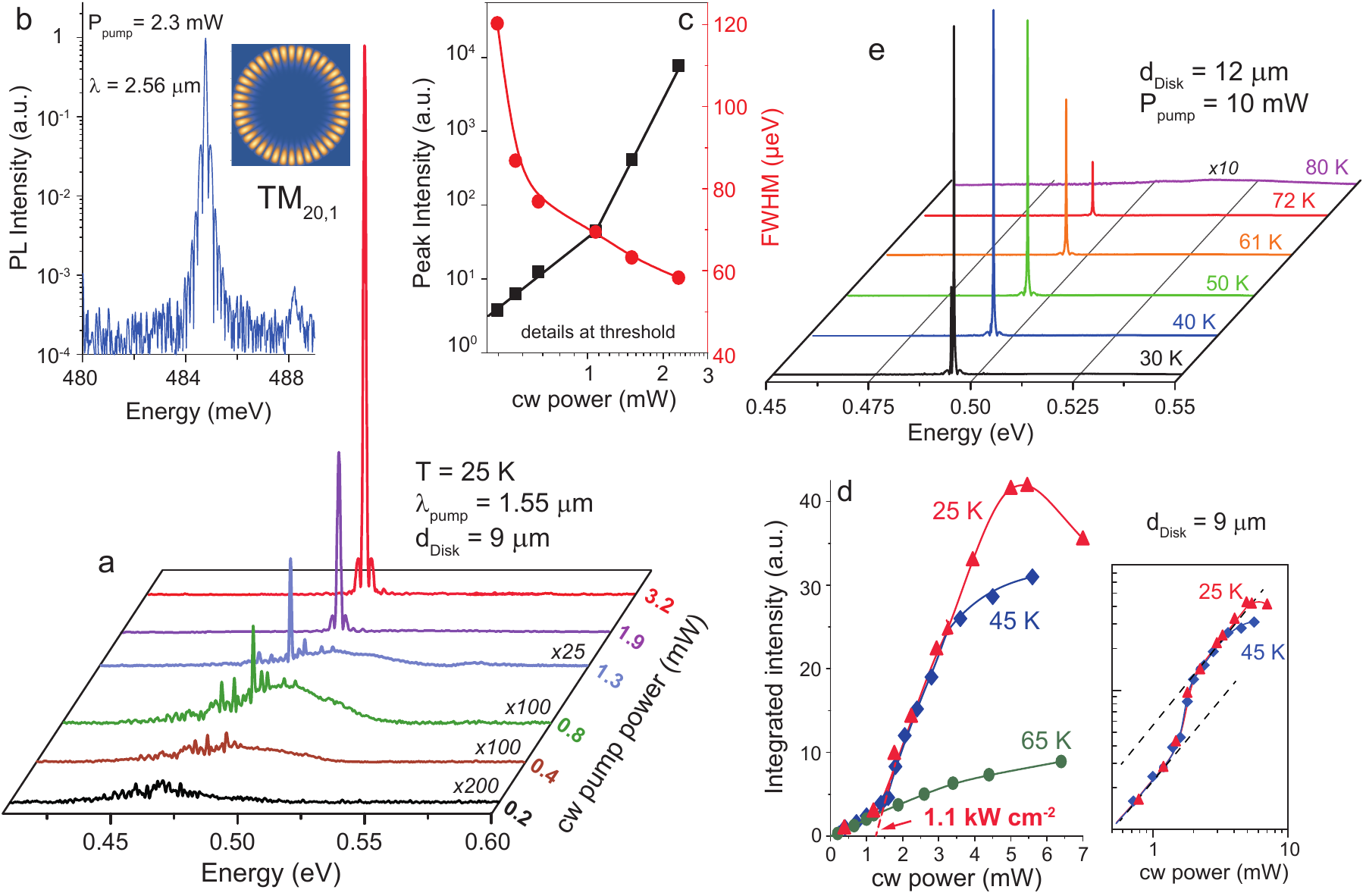}
\caption{\textbf{CW lasing from GeSn.} \textbf{(a)} Emission spectra measured at \SI{25}{\kelvin} for the \SI{9}{\micro\meter} diameter disk under various cw excitation levels. \textbf{(b)} Lasing mode peak intensity above threshold in log scale, highlighting its high intensity compared to the spontaneous background emission. The intensity profile of the lasing mode, identified as TM$_{20,1}$, is shown in the inset. \textbf{(c)} Detailed behaviour at threshold: L-L curve of the peak intensity of the \SI{485}{\milli\electronvolt} laser mode (black) and its linewidth (red). \textbf{(d)} L-L curve of the integrated spectra at different temperatures (on logarithmic scales in the inset). Lasing threshold is extracted to be \SI{1.3}{\milli\watt} (see SI), corresponding to \SI{1.1}{\kilo\watt\per\square\centi\meter}. \textbf{(e)} PL spectra measured on a \SI{12}{\micro\meter} diameter disk under cw pump power of \SI{10}{\milli\watt} at various temperatures from \SIrange{30}{80}{\kelvin} showing lasing operation up to \SI{72}{\kelvin}.}
\label{fig:Lase}
\end{figure}

 In order to obtain stimulated emission, microdisk devices with a diameter of \SI{9}{\micro\meter} were optically pumped using a $\mu$-PL setup with \SI{1550}{\nano\meter} wavelength cw pump laser focused on the sample surface into a \SI{12}{\micro\meter} diameter spot (see Methods). PL emission spectra collected at various incident pump powers at \SI{25}{\kelvin} are shown in Figure~\ref{fig:Lase}a. At low excitation levels the microdisks produce a broad spontaneous emission background, attributed to $\Gamma\to LH$ direct transitions. By increasing the cw pump power from \SI{0.2}{\milli\watt} to \SI{0.8}{\milli\watt}, whispering gallery modes (WGM) develop and grow in intensity on top of the spontaneous emission. Higher excitation induces an exponential intensity increase of the main optical mode at \SI{485}{\milli\electronvolt}. At \SI{2.3}{\milli\watt} pump power the lasing emission is four orders of magnitude stronger than the background, as shown in the high resolution spectrum in Figure~\ref{fig:Lase}b. The two symmetric side lobes of the modes are only a measurement artifact, which stems from a finite range of sampling points due to the apodization of the interferogram.\cite{Rabolt1981}\\
The observation of a clear threshold in the light-in light-out (L-L) characteristic visible in Figure~\ref{fig:Lase} c,d, the $S$-shape L-L characteristic (inset Figure~\ref{fig:Lase}d) and the collapse of the linewidth (Figure~\ref{fig:Lase} c) unambiguously prove the onset of lasing. The emission energy at \SI{25}{\kelvin} of the lasing mode at \SI{485}{\milli\electronvolt} corresponds to \SI{2.55}{\micro\meter} wavelength. WGM simulations (see Methods) indicate that this mode is the transverse magnetic TM$_{20,1}$ mode. The intensity of this mode in the threshold region is displayed in Figure~\ref{fig:Lase}c as a function of incident pump power by the linear L-L characteristics. The observed narrowing is consistent with the Schawlow-Townes equations, which predict a decrease by a factor of two at the transition from incoherent to coherent emission. Note that the measured linewidth of \SI{58}{\micro\electronvolt} is the smallest reported to date for any group-IV semiconductor laser.\\
The laser threshold, clearly separating the spontaneous and the amplified emission regimes, is extracted as \SI{1.3}{\milli\watt} (Figure~\ref{fig:Lase}d). This value corresponds to a pump power density of \SI{1.1}{\kilo\watt\per\square\centi\meter}. The dependence of the integrated signal on cw pumping power at different temperatures is also shown in Fig.\ref{fig:Lase}d. The typical laser S-shape emission is shown in logarithmic scales in the inset. Below \SI{45}{\kelvin} an unambiguous threshold can be observed, while no signature of lasing is detected above \SI{45}{\kelvin}. A roll-over of the integrated emission occurs for pump powers above \SI{5}{\milli\watt}. In III-V compounds this effect is typically associated with thermal effects, leading to a sharp laser emission quenching with further increase of temperature. However, in the case of GeSn microdisk laser under investigation, the directness $\Delta E_{L-\Gamma}$ is in the range of only $\SI{70}{\milli\electronvolt}$. In fact, band filling effects of the $\Gamma$-valley will reduce the electrons energy required for thermal (scattering) escape from $\Gamma$- into the $L$-valley even further. Therefore, a small increase in temperature or excitation power may lead to an exponential increase of this thermal escape, producing the observed roll-over. As will be discussed later, an increase in the electron population of the $L$-valley leads to a decrease of the total gain. Microdisks with diameter of \SI{12}{\micro\meter} were also fabricated. The undereteching is kept constant leading to same biaxial tensile strain in the GeSn suspended area, where the WGMs are formed. The laser emission at \SI{489}{\milli\electronvolt}, is only \SI{4}{\milli\electronvolt} higher than in  \SI{9}{\micro\meter} diameter disks. One major difference here is that the Al support pillar is much larger now, enabling a better heat dissipation. The \SI{12}{\micro\meter} diameter disk can support the cw lasing up to \SI{72}{\kelvin}. The PL spectra versus temperature are plotted in Figure~\ref{fig:Lase}e. The same laser threshold is obtained as for the smaller \SI{9}{\micro\meter} diameter disks and increases with temperature reaching about \SI{5}{\milli\watt} at \SI{72}{\kelvin} (details in SI).
Estimation of the disk heating under optical pumping, using Finite Element Modeling (FEM) \cite{Elbaz:18, el_kurdi_tensile-strained_2016}, can be found in the SI.

\begin{figure}[ht]
\centering
\includegraphics[width=0.5\textwidth]{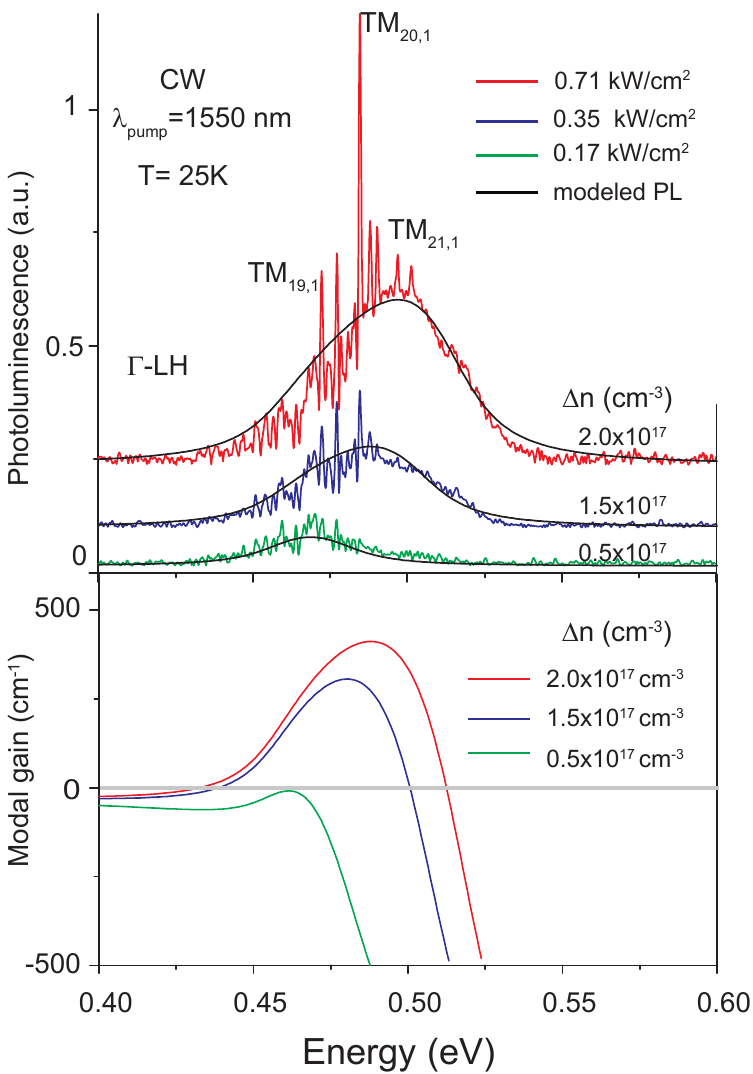}
\caption{\textbf{Carrier density at threshold and net gain} \textbf{(a)} PL spectra below threshold, taken at \SI{25}{\kelvin} under cw pumping of \SI{0.2}{\milli\watt}, \SI{0.4}{\milli\watt} and \SI{0.8}{\milli\watt}. WGMs form on top of the spontaneous emission-dominated spectrum. The solid black lines are the calculated spontaneous emission spectra for different carrier densities.\textbf{(b)} Calculated modal gain for the carrier density values from (a).}
\label{fig:Gain}
\end{figure}

The PL spectra of a \SI{9}{\micro\meter} diameter disk taken at excitation powers below the lasing threshold are shown in Figure~\ref{fig:Gain}a. Assuming a constant temperature of \SI{25}{\kelvin}, spontaneous emission spectra were calculated for different carrier densities in order to reproduce the broadening of the emission and its blue shift when the pump power increases. The emission blue shift, observed experimentally when increasing the pump power, can be attributed to band filling effect inducing transitions at higher energies, since additional optical transitions occur at higher energies. Considering a band-gap of \SI{465}{\milli\electronvolt} the broadened PL spectra are well reproduced for carrier densities of \SI{0.5d17}{\per\cubic\centi\meter}, \SI{1.5d17}{\per\cubic\centi\meter} and \SI{2d17}{\per\cubic\centi\meter}, which correspond to pump intensities of \SI{0.17}{\kilo\watt\per\square\centi\meter}, \SI{0.35}{\kilo\watt\per\square\centi\meter} and \SI{0.71}{\kilo\watt\per\square\centi\meter}, respectively.\\ 
Under cw pumping, the steady state carrier density can be directly linked to the carrier lifetime, using the generation-recombination balance law $N=I\tau/(h\nu d)$, where $I$ is the absorbed power density, $\tau$ the non-radiative recombination lifetime, $h\nu$ the pump photon energy and $d$ the absorption depth. To obtain the absorbed power used for calculating carrier density and material gain, the incident power is multiplied with factor of 0.65 which accounts for the disk surface reflectivity at \SI{1550}{\nano\meter} wavelength. The extracted non-radiative lifetimes are \SI{1.4}{\nano\second}, \SI{2.1}{\nano\second} and \SI{1.4}{\nano\second} for carrier densities of \SI{0.5d17}{\per\cubic\centi\meter}, \SI{1.5d17}{\per\cubic\centi\meter} and \SI{2d17}{\per\cubic\centi\meter}, respectively. The obtained values, averaged to around \SI{1.6}{\nano\second}, are equivalent to those reported in Ref.~\cite{wirths_lasing_2015}. However, even with comparable lifetimes, a clearly lower pumping threshold is here observed. This remarkable observation will be discussed in the next section. \\
The modal gain illustrated in Figure~\ref{fig:Gain}b was calculated for the carrier densities mentioned above, assuming steady-state conditions. The required parameters were obtained from the \textbf{k$\cdot$p}  band structure description, and are summarized in Methods. At a carrier density of \SI{0.5d17}{\per\cubic\centi\meter} the positive gain regime is not reached, but the gain steeply increases with the carrier density. The modal gain maximum is observed at the lasing mode energy of \SI{485}{\milli\electronvolt}. Furthermore, gain broadening is observed for larger pumping intensity due to band filling effects. Note that the calculated modal gain might be overestimated, since it depends on unknown parameters such as homogeneous broadening. Nonetheless, these calculations indicate that modal gain can reach up to \SI{400}{\per\centi\meter} for carrier densities of \SI{2d17}{\per\cubic\centi\meter}.\\
Narrow whispering gallery modes appear in addition to the broad spectrum of spontaneous emission, clearly visible in Figure~\ref{fig:Gain}a. The modes at \SI{472}{\milli\electronvolt}, \SI{485}{\milli\electronvolt} and \SI{497}{\milli\electronvolt} are regularly spaced by \SIrange{12}{13}{\milli\electronvolt} and can be assigned to fundamental TM$_{19,1}$ TM$_{20,1}$ and TM$_{21,1}$ modes, respectively. Other mode patterns in the spectrum can be attributed to higher radial index modes, with $n = 2$ and $n = 3$. The TM$_{19,1}$ mode is close to the band-gap, just a few \si{\milli\electronvolt} apart. According to gain modelling it does not match the gain maximum. The TM$_{20,1}$ mode, at higher energy, here shows a better match. Thus the TM$_{20,1}$ mode will dominate the spectrum at high powers. This mode competition behaviour leads to a clear single-mode laser emission.  

\begin{figure}[ht]
\centering
\includegraphics[width=\linewidth]{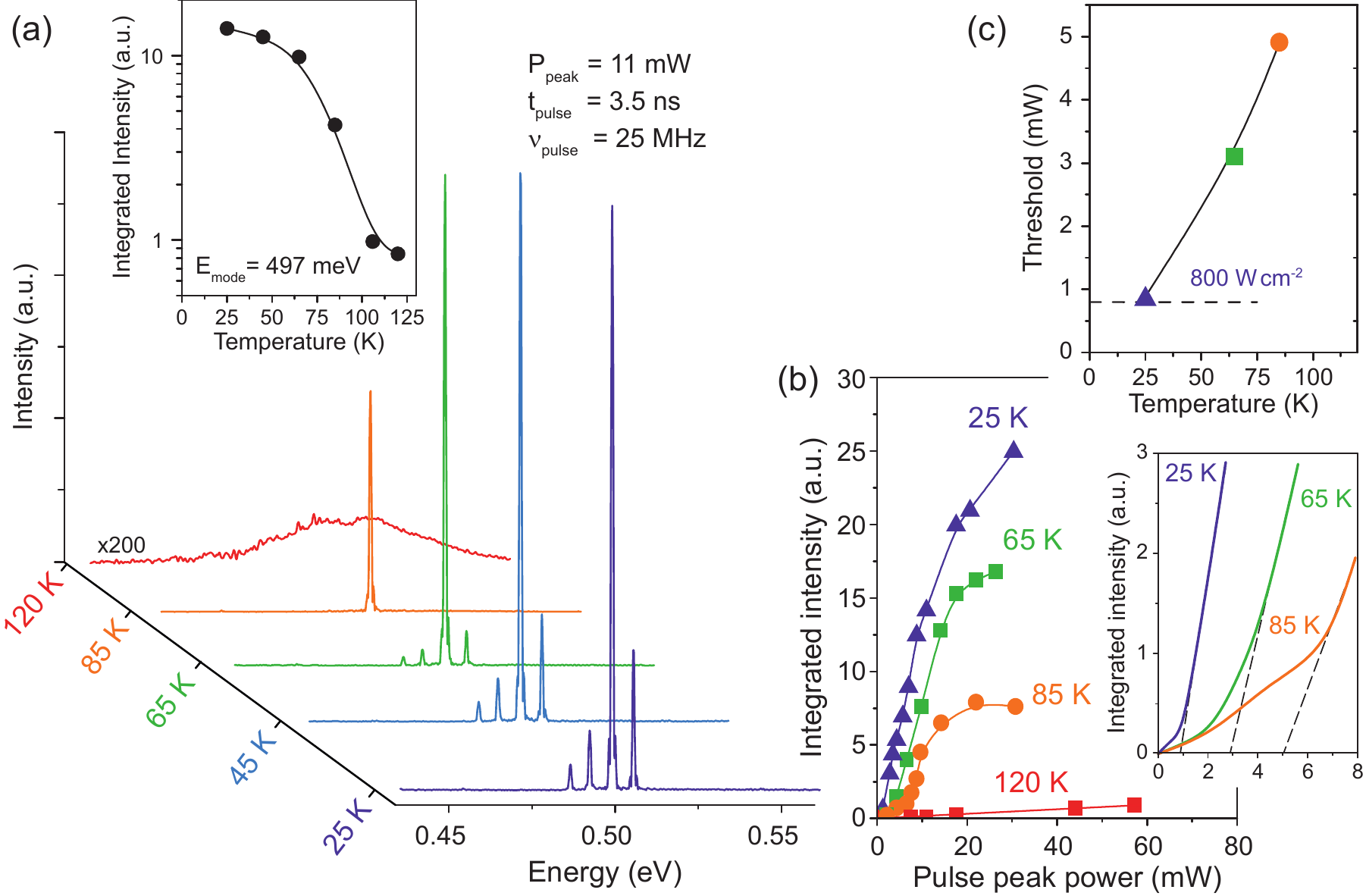}
\caption{ \textbf{Pulsed laser.} \textbf{a} Temperature dependence of light emission of a \SI{9}{\micro\meter} diameter disk under pulsed peak power excitation of \SI{11}{\milli\watt}. The inset shows the integrated PL signal for the laser mode at \SI{497}{\milli\electronvolt}. \textbf{b} L-L curves at \SI{25}{\kelvin}, \SI{65}{\kelvin}, \SI{80}{\kelvin} and \SI{120}{\kelvin}. No lasing is observed at \SI{120}{\kelvin}. (inset) Zoomed view of L-L curves around threshold. \textbf{c} Lasing threshold as a function of temperature.} 
\label{fig:LasePulse}
\end{figure}

 Since all the GeSn lasers reported in the literature were only operational under pulsed laser excitation, we also studied the properties of the present tensile strained GeSn microdisks. This enables a comparison of GeSn lasers with moderate Sn contents and tensile strain to those reported previously in the literature, having high Sn contents and residual compressive strain. Laser emission spectra taken at temperatures in the range of \SIrange{25}{120}{\kelvin} are shown in Figure~\ref{fig:LasePulse}a.  \SI{3.5}{\nano\second} long pulsed excitation at a fixed peak pump power of \SI{11}{\milli\watt} (\SI{1}{\milli\watt} average pump power) was here applied. Distinct differences from the cw excitation regime are observed. The laser spectra are multi-mode, with the dominant mode at \SI{497}{\milli\electronvolt}, corresponding to \SI{2.49}{\micro\meter} emission wavelength. Mode assignment can be made by referencing to the spectra shown in Figure~\ref{fig:Gain}a. The lasing mode at the lowest energy corresponds to the one under cw operation at \SI{485}{\milli\electronvolt}. The dominant mode at \SI{497}{\milli\electronvolt} is also observed at the same energy in Figure~\ref{fig:Gain}a. From modelling of the whispering gallery modes, we can assign this mode to the transverse magnetic TM$_{21,1}$, as the free spectral range matches the expectation from modelling. The other two modes, at \SI{488}{\milli\electronvolt} and \SI{502}{\milli\electronvolt}, are assigned to modes with a higher order radial index, i.e. TM$_{17,2}$ and TM$_{18,2}$. The modes with radial number $n = 1$ are essentially un-shifted compared to those observed in cw regime in Figure~\ref{fig:Gain}a. Under pulsed excitation, there is a non-uniform spatial distribution of carriers, and possibly also a non-equilibrium distribution over the conduction band valleys, as well as a different temperature gradient compared to cw excitation. The non-uniform/non-equilibrium distribution explains the shift of gain to modes with higher radial order indices. In the pulsed regime, the gain profile may indeed be different from that in cw, enabling multi-mode lasing and favouring higher energy modes. At higher temperatures, the laser spectra become single-mode (Figure~\ref{fig:LasePulse}a), indicating a narrower frequency range of positive gain. However, more importantly, the maximum laser operation temperature in pulsed operation is almost two times higher than in cw regime.\\
The L-L characteristics for different temperatures are given in Figure~\ref{fig:LasePulse}b. Here, a thermal roll-over occurs at \SI{85}{\kelvin} for \SI{20}{\milli\watt} peak pumping power (\SI{1.8}{\milli\watt} average power), compared to \SI{5}{\milli\watt} at \SI{25}{\kelvin} under cw excitation.  For pulsed operation at \SI{25}{\kelvin} a lasing threshold peak power of \SI{0.9}{\milli\watt} is measured (inset of Figure~\ref{fig:LasePulse}b), which corresponds to a record low pump power density of \SI{0.8}{\kilo\watt\per\square\centi\meter}. This pump power density is 250 times smaller than that of the best bulk GeSn microdisk laser reported in the literature. The threshold is temperature dependent, as shown in Figure~\ref{fig:LasePulse}c. A rise in temperature to \SI{85}{\kelvin} increases the threshold power density to \SI{4.2}{\kilo\watt\per\square\centi\meter}. At \SI{120}{\kelvin} no lasing signature is observed anymore. The maximum temperature at which the gain should be still detectable depends on the carriers density as well. Simulations of the gain for an electron / hole density of \SI{1d17}{\per\cubic\centi\meter} indicate that the gain in this material becomes quite small at \SI{120}{\kelvin}. Taking parasitic losses, like scattering on surface roughness into account,  these simulation results match the experimental observation.\\
In this paragraph we qualitatively discuss the laser operation mode, maximum lasing temperature and threshold pump power density for GeSn layers under tensile strain, and benchmark the results against literature data. \\
In our experiment, where pumping is above the lasing threshold, the steady-state carrier density is expected to be nearly constant since the quasi- Fermi level is clamped by stimulated emission process. Although all electrons are generated in the $\Gamma$-valley, many of them will scatter into the $L$-valley. In particular, for a small $\Delta E_{L-\Gamma}$, here only \SI{70}{\milli\electronvolt}, and an energy difference between the quasi Fermi level and the $L$-valley of only \SI{40}{\milli\electronvolt}, estimated at 25K for an electron-hole density of \SI{2d17}{\per\cubic\centi\meter}, a substantial amount of carriers will cool down/scatter into the $L$-valley. The intra- and inter-valley carrier scattering times, by phonon emission/absorption, are in the \si{\pico\second} range while the thermal relaxation time for the disk is in the \si{\micro\second} range. This implies an equilibrium carrier distribution in cw, at a lattice temperature above the nominal temperature, by about \SIrange{20}{30}{\kelvin} under \SI{8}{\milli\watt} pump power (see SI).  The scattering rates increase with temperature and with the decrease of $\Delta E_{L-\Gamma}$, increasing the carrier density in the $L$-valleys superlinearly with the pump power. This leads to  a decrease of the gain since carriers in $L$-valleys only contribute to free carrier absorption without any contribution to gain.  Moreover, as the temperature is increased, the carrier density is affected by increased non-radiative recombinations that decreases the overall gain (see SI). Consequently, a roll-over of the L-L spectrum at \SI{25}{\kelvin} is observed for an excitation power exceeding \SI{7}{\milli\watt} and above \SI{45}{\kelvin} the lasing is strongly quenched. These results hint to a heating of the electron bath by \SI{40}{\kelvin} above the nominal sample temperature. It was experimentally observed that reducing the pump photon energy, and consequently reducing the lattice heating, leads to increased maximum operating temperatures by about \SI{20}{\kelvin}  \cite{Dou2018a,StangeMQWlaser}.
The role of the lattice heating is underlined by the experimental observation of operation temperature increase by almost \SI{30}{\kelvin} for the \SI{12}{\micro\meter} disks, because the heat disposal via Al pillar there is better than for the \SI{9}{\micro\meter} disks. \\
Operating the same laser under pulsed excitation substantially reduces the thermal load, improving the temperature performance of the device. Laser quenching appears above \SI{85}{\kelvin} for \SI{9}{\micro\meter} disks and \SI{100}{\kelvin} for \SI{12}{\micro\meter} disks. The operation temperature increase for pulsed operation is only \SI{15}{\kelvin}, about a half of that for cw operation. This indicates that cw lasing is much more sensitive to temperature than the pulsed operation. This may explain why no cw lasing was observed in compressively strained GeSn laser where the pumping powers were 100-200 times larger than in this work. Multi-mode and single-mode lasing for pulsed and cw pumping, respectively, is likely related to an incomplete wash-out of spatial hole-burning by carrier diffusion, together with a larger gain, for pulsed pumping \cite{sargent_1993}.\\
Next we compare the results with those for Ge$_{0.875}$Sn$_{0.125}$ microdisk laser with residual compressive strain, with a similar diameter of \SI{8}{\micro\meter} \cite{stange_optically_2016}, emitting at almost the same energy of \SI{\sim 0.5}{\electronvolt} (\SI{2.5}{\micro\meter}), having similar band-gap and directness $\Delta E_{L-\Gamma}$. With much stronger pump pulses, but 3 orders of magnitude smaller duty cycle, the average pump power and heat load there were lower, so it could tolerate a larger gain drop with increasing temperature, enabling a larger maximum operating temperature, of  $\sim$\SI{130}{\kelvin}. The laser active layer in this work is the thinnest reported to date, less than a half of that in Ref.\cite{stange_optically_2016} (\SI{260}{\nano\meter} vs. \SI{560}{\nano\meter}), hence the mode overlap factors (affecting modal gain) also differ, being just \SI{17}{\percent} here and \SI{95}{\percent} in Ref.\cite{stange_optically_2016}. Even so, the GeSn device strained by SiN$_{x}$ layers shows substantially lower threshold. \\
The main achievement of the presented GeSn laser is that it enables the cw operation. This is a direct consequence of the strongly reduced threshold power, by at least a factor of \SI{250}, which allows continuous pumping while maintaining a moderate lattice temperature. This drastic improvement is attributed to the three factors: {\em (i)} The large reduction of the valence band DOS ($\Delta$E$_{LH-HH}=$ \SI{172}{\milli\electronvolt}), warrants that the population inversion is achieved at a lower density of excited carriers. {\em (ii)} The modal gain of tensile strained GeSn is improved compared to unstrained GeSn with higher Sn content and the same $\Delta$E$_{L-\Gamma}$ \cite{Rainko2018a}. {\em (iii)} The non-radiative recombination rate is substantially reduced in lower Sn content alloys and even more so in the present structure with the defective epitaxial interface intentionally removed during processing. Defect reduction engineering is crucial, as previously reported for GeSn/SiGeSn quantum wells (QWs) laser \cite{StangeMQWlaser}. \\ 
The above discussion should provide some guidelines for improving the laser performance, by combining the advantages of the present structure with those from the previous GeSn laser research. An increase of the operating temperature up to \SI{230}{\kelvin} was obtained by increasing the Sn content to \SI{16}{\atomic\percent}\cite{Thai:18}, and hence the directness $\Delta$E$_{L-\Gamma}$. Tensile strain can considerably reduce the requirement for high Sn contents. A tensile strain of \SI{1.5}{\percent} in \SI{10}{\atomic\percent} Sn alloy results in a very large $\Gamma$ population at \SI{300}{\kelvin} (see SI), and is, therefore, a viable route towards application-ready GeSn lasers. In addition, the use of SiGeSn/GeSn QWs heterostructures brings the additional benefits from carrier confinement and energy quantization enabling a threshold reduction by almost an order of magnitude for strain relaxed SiGeSn/GeSn QW \cite{StangeMQWlaser}. Heterostructures like presented in Ref.\cite{Stange_MQW_LED}, combined with tensile strain, would strongly improve the carrier confinement at higher temperatures and, consequently, further decrease the threshold pump power density. QWs heterostructures design allows the separation of the active region from interface defects without using a layer transfer technology, which is desirable for low-cost, high yield fabrication. In the present work bonding was used as a straightforward way to identify the intrinsic material optical properties. For tensile strain engineering, however, layer transfer is also not a mandatory solution, since several previous works have shown the feasibility of using a single stressor in cavities within a fully monolithic approach \cite{capellini_strain_2013,ghrib_tensile-strained_2013,ghrib_control_2012}. Finally, the use of stessor layers has been shown to be compatible with electrical pumping scheme \cite{prost_tensile-strained_2015} that could be incorporated into the presently used layout.

\section*{Conclusion}
In summary, we have demonstrated a GeSn-based laser fully embedded in a SiN$_{x}$ stressor conform layer. Starting with an indirect band-gap GeSn alloy we converted it, by inducing tensile strain, into a direct-gap optical gain material. Most importantly, cw and pulsed laser operation with ultra-low lasing thresholds of \SIrange{0.8}{1.1}{\kilo\watt\per\square\centi\meter} are demonstrated. These values are significantly lower than any value reported for group-IV lasers, and are even comparable to those for epitaxially grown III-V InP laser \cite{Wang2017a}, or InGaAs lasers bonded on Si wafers\cite{Seifried2018}, albeit the latter operate at room temperature. This achievement relies on using dilute GeSn alloys with just \SI{5.4}{\atomic\percent} Sn, in contrast to the mainstream research focusing on \SIrange{16}{20}{\atomic\percent} Sn. While the lower Sn content decreases the $\Delta E_{L-\Gamma}$ splitting (directness), tensile strain compensates for that and, additionally, offers a reduced DOS by shifting the $LH$ band above the $HH$ band. In the present work, we combine the material quality advantage of low Sn content alloys and the physics of tensile strain with processing technology and thermal management, enabling lasing in low Sn-content alloys at record-low pumping powers and operating in a cw regime.

\section*{Methods}

Optical PL excitation was provided by an Nd:YAG laser at \SI{1064}{\nano\meter} wavelength focused to the sample surface with a microscope objective, $\times$40 and NA of 0.65, into a \SI{5}{\micro\meter} diameter spot. Collection of emitted light was performed with the same microscope objective, and a beamsplitter was used to separate emission and excitation beam paths. In these experiments, we used an extended InGaAs detector to detect the luminescence. The high sensitivity, enabling the detection of low levels of infrared signals, was counterbalanced by the \SI{2.5}{\micro\meter} wavelength cut-off due to the detector and objective transmission quenching.
\newline
\newline
Lasing experiments were performed using a $\mu$-PL setup, where the cw pump laser beam at \SI{1550}{\nano\meter} wavelength was focused on the sample surface into a \SI{12}{\micro\meter} spot diameter by a x40 reflective objective with numerical aperture of 0.5 and working distance of \SI{4}{\milli\meter}. The same objective was used for pumping and for collection of light emitted from the microdisk. A CaF$_2$ beamsplitter was used to separate the excitation and emission beam paths. The outgoing emission, collected from the objective, was coupled to a Fourier Transform Infrared (FTIR) spectrometer equipped with a CaF$_2$ beam splitter. The emission was detected by a nitrogen-cooled InSb photodetector, which has a cut-off wavelength \SI{4.8}{\micro\meter}. The telecom wavelength pump laser was out-coupled from a single-mode fiber to free space using a hyperbolic mirror, the output fiber being clamped at the focal point of the mirror. The pump can be taken through a Mach-Zehnder modulator, controlled by an rf-pulse generator so that the pump beam power can be switched from cw to quasi-continuous and to pulsed mode. The same output from a single-mode fiber was used in both cases, cw or pulsed, so that switching from one pump mode to another did not induce any change of the beam waist or its alignment to the disk. The pulse shape under modulated excitation had \SI{3.5}{\nano\second} width and a repetition rate of \SI{25}{\mega\hertz}. As discussed below, the pulse duration was longer than the non-radiative lifetime, so the optical excitation can be considered quasi-cw in this respect. 
\newline
\newline
Calculation of whispering gallery modes was performed by a 2D analytical model. The resonance wavelength of the cavity mode with azimuthal index m is calculated from the roots of m\textsuperscript{th} Bessel function $I_{\text{m}} (\frac{2\pi n_{eff}(\lambda)}{\lambda_{res}}a)$. The optical field is plotted at the resonant wavelength to obtain the corresponding radial number n of nodes along the disk radius. To account for modal dispersion of vertically confined modes, the effective index is introduced in the model, according to $n_{eff}(\lambda)=-0.6\lambda +3.85$, where $\lambda$ is the wavelength in \si{\micro\meter}. This expression was interpolated from 1D slab modelling of TM polarized waves, propagating in the GeSn layer. The energy of the TM$_{20,1}$ mode is calculated to be \SI{484.65}{\milli\electronvolt}, in good agreement with the \SI{485}{\milli\electronvolt} experimental value.
\newline
\newline
Optical gain is calculated within the framework of parabolic band effective mass model, using the equation \cite{ghrib_all-around_2015} 
\begin{equation*}
g(\hbar\omega)= \left|D_{\Gamma,LH}^{\text{TM}}\right|^2 \int\limits_{0}^{\infty} \rho_{LH}(E)(f_C(E)-f_V(E)) \frac{\frac{\Gamma_0}{2\pi}}{(\frac{\Gamma_0}{2})^2 + (E_{\Gamma,LH} + E-\hbar\omega)^2} \, \mathrm{d}E,
\end{equation*}
where $\rho_{LH}(E)$ is the joint density of states involving the $\Gamma$ conduction band and the $LH$ valence band, and $\left|D_{\Gamma,\text{LH}}^{\text{TM}}\right|^2$ is the interband dipole matrix element. $\Gamma_0$ is the full width at half maximum of a Lorentzian function, which is here set to \SI{25}{\milli\electronvolt} to account for homogeneous broadening of interband transitions, while $f_C(E)$ and $f_V(E)$ are the Fermi-Dirac functions for electrons and holes. The density of states effective masses, used in the calculation, were $m_\Gamma=0.036~m_0$, $m_L=0.56~m_0$, and $m_{LH}=0.05~m_0$, extracted from the \textbf{k.p} model\cite{Rainko2018a} of strained GeSn with \SI{5.4}{\atomic\percent} Sn content and biaxial tensile strain of \SI{1.4}{\percent}. The energy splitting $\Delta E_{L-\Gamma} = E_L-E_\Gamma$, $\Delta E_{LH-HH} = E_{LH}-E_{HH}$ and the band-gap $E_\Gamma-E_{LH}$ amount to \SI{70}{\milli\electronvolt}, \SI{170}{\milli\electronvolt} and \SI{465}{\milli\electronvolt},  respectively, according to \textbf{k.p} modeling and PL results. Note that the density of states effective masses differ only slightly from those in pure Ge, and we have therefore used the free carrier absorption model for pure Ge, assuming it would be very close to that for low Sn content GeSn alloy, in order to deduct it from the interband gain and find the net gain.

\section*{Acknowledgements }
M.E.K. and A.E. thank Dr. Raffaele Colombelli and Prof. Adel Bousseksou for fruitful discussions and their help in mounting the PL setup with the FTIR spectrometer. The authors thank Dr. Gregor Mussler for XRD measurements. This work used knowledge acquired in the collaboration with Dr. Hans Sigg from PSI. This work was supported by the French RENATECH network, the French national research agency (Agence Nationale de la Recherche - ANR) through funding of ELEGANTE project (ANR-17-CE24-0015) and the Deutsche Forschungsgemeinschaft (DFG) via the project “SiGeSn Laser for Silicon Photonics”. A.E. was supported by ANRT through a CIFRE grant.  A. F. gratefully acknowledges funding within the ANR-16-CE09-0029-03 TIPTOP project.

\section*{Author contributions statement}

All authors contributed to the work. P.B. M.E.K. and A.E. designed the device structure. M.E.K. and A.E. performed the strained disks fabrication with E.H., I.S., K.P., and G.P.. M.E.K. and A.E. performed the PL measurements and laser experiments with N.Z. and X.C.. K.P., G.P., I.S., N.v.d.D. and D.B. performed the structural analysis of the material. The GeSn layer was grown by D.B. and N.v.d.D. on substrates from J.-M.H.. The Raman analyses were performed by A.F. and R.O.. P.B., S.S. and Z.I. contributed to the modelling with M.E.K. and D.B.. The work was supervised by D.G., F.B., P.B., D.B. and M.E.K.. P.B, M.E.K., N.v.d.D., D.G and D.B. wrote the manuscript.
\newline

\section*{Competing interest}
The authors declare no competing financial interests.
\newline
\section*{Data availability statement}
The authors declare that all the data supporting the findings of this study are available within the paper and its supplementary information file. 
\section*{Code availability statement}
Finite element modeling was performed using a commercially available COMSOL software. All-other calculation codes were used in published works where model details are provided. The codes are not publicly available, any requests should be sent to the corresponding authors.

\end{document}